\documentclass[sn-mathphys-ay]{sn-jnl}

\usepackage{graphicx}%
\usepackage{multirow}%
\usepackage{amsmath,amssymb,amsfonts}%
\usepackage{amsthm}%
\usepackage{mathrsfs}%
\usepackage[title]{appendix}%
\usepackage{xcolor}%
\usepackage{textcomp}%
\usepackage{manyfoot}%
\usepackage{booktabs}%
\usepackage{algorithm}%
\usepackage{algorithmicx}%
\usepackage{algpseudocode}%
\usepackage{listings}%
\usepackage{tikz}

\theoremstyle{thmstyleone}%
\theoremstyle{thmstyletwo}%

\theoremstyle{thmstylethree}%

\setcitestyle{uniquename=false}

\begin{document}

\title[Article Title]{Backward similarity solution of the Boussinesq groundwater equation}


\author*[1]{\fnm{Shuntaro} \sur{Togo}}\email{togo.shuntaro.66s@st.kyoto-u.ac.jp}

\author[1]{\fnm{Koichi} \sur{Unami}}\email{unami.koichi.6v@kyoto-u.ac.jp}

\affil[1]{\orgdiv{Graduate School of Agriculture}, \orgname{Kyoto University}, \orgaddress{\city{Kyoto}, \postcode{606-8502}, \country{Japan}}}


\abstract{Groundwater flow in an unconfined aquifer resting on a horizontal impermeable layer with a boundary condition of a rapid increase in the source water level is considered in this work. The newly introduced condition, referred to as the backward power-law head condition, represents a situation where the water level in the adjacent water body increases more rapidly than do conventional problems, which can only represent a situation akin to a traveling wave under rising water level conditions, given its consideration of infinite time. This problem admits the similarity transformation which allows the nonlinear partial differential equation to be converted into a nonlinear ordinary differential equation via the Boltzmann transformation. The reduced boundary value problem is interpreted as the initial value problem for a system of ordinary differential equations (ODE), which can be numerically solved via Shampine's method. The numerical solutions are in good agreement with Kalashinikov's special solution, which is also introduced into the Boussinesq equation. We find that the solution is consistent with the limit of the forward power-law head condition. The new approximate analytical solution and the associated wetting front position are derived by assuming that the solution has the form of quadratic polynomials, which enables the description of the time progression of a real front position. The obtained approximation is compared to Shampine's solution to check the accuracy. Furthermore, the finite element method is applied to the original partial differential equation (PDE), which validates Shampine's solution for use as a benchmark.}

\keywords{Groundwater, Boussinesq equation, Quadratic approximate solution, Shampine's method}



\maketitle

\section*{Article Highlights}
\begin{itemize}
    \item We propose numerical and approximate solutions for a newly introduced rapidly increasing water level condition.
    \item The derived similarity solutions visualize diverse water level rise scenarios beyond previous models.
    \item We compare numerical solutions from similarity calculations with direct numerical computations.
\end{itemize}

\section*{Acknowledgement}
This work was supported by Grant-in-Aid for Scientific Research No.19KK0167 from the Japan Society for the Promotion of Science (JSPS), Research Promotion Support 2024A from the Kyoto University Foundation.

\section{Introduction}
\label{intro}

Unconfined groundwater flow is usually modeled by the Boussinesq groundwater equation under Dupuit's assumption that the groundwater flows horizontally in the direction of the surface gradient. At steady state,  the governing equation is regarded as the well-known Laplace equation since the partial derivative term with respect to time is equal to zero. On the other hand, in the nonstationary case, the derivative remains and the equation is considered to be a nonlinear diffusion equation, known as the porous media equation ($m = 2$) :
\begin{equation}
	\frac{\partial u}{\partial t} = \nabla \cdot \nabla u^m = \nabla \cdot \left( mu^{m-1}\nabla u \right), \label{eq:PME}
\end{equation}
refer to \citet*{Vazquez2007, Gilding1982, Furtak-Cole2018, Kacimov2021}. \citet{Gilding1982} classified the similarity solutions of the one-dimensional porous media equation (\ref{eq:PME}) into three groups depending on its boundary conditions, the forward condition, the backward condition, and the exponential condition, which enables the partial differential equation to be changed into the ordinary differential equation. It is also shown that all the known explicit analytical solutions are similarity solutions. 
These three types of boundary conditions are applied to groundwater hydrology as the forward power law head condition \citep*{Barenblatt1990,Chen1995,Lockington2000,Telyakovskiy2002, Telyakovskiy2006, Song2007, Furtak-Cole2018}, the forward power-law flux condition \citep*{Barenblatt1990, Telyakovskiy2006}, the backward power-law flux condition \citep{Barenblatt1990}, the exponential power-law head condition \citep*{Barenblatt1990,Telyakovskiy2006}, and the exponential power-law flux condition \citep{Telyakovskiy2006} for an initially dry unconfied aquifer, which represents the water-level fluctuations of an adjacent water body as time passes.  In particular, the forward power law head condition, if the exponent is zero, corresponds to the constant water head condition describing sudden water change, which has been widely studied \citep*{Polubarinova1962, Tolikas1984, Chor2013, Chor2019, Hayek2024}. For the initially saturated aquifer, the constant inlet condition addresses the drainage problem, of which a detailed explanation can be found in \citet{Rupp2005} who applies the intrinsic permeability proportional to depth to the recession analysis first proposed by \citet{Brutsaert1977}. 

 Furthermore, many explicit special solutions of the porous media equation written in \citep{Gilding1982} are also considered groundwater flow solutions, such as the dipole solution \citep*{Barenblatt1990, Vazquez2007}, the point-source solution \citep{Barenblatt1990, Kacimov2021}, and the traveling-wave solution \citep*{Barenblatt1990, Chen1995}. We recall that the latter two solutions are referred to as Barenblatt's exact solutions to validate the approximate solutions in some studies \citep*{Chen1995, Parlange2000, Lockington2000}. 

To address the difficulty of obtaining the general solution of the Boussinesq equation, numerous approximate analytical solutions have been presented for simple geometries and boundary conditions, which have the advantages of examining the validity of the numerical schemes and facilitating the understanding of physical phenomena such as the finite propagation of the waterfront and scaling properties \citep{Telyakovskiy2006}. \citet{Serrano1998} applied Adomian's decomposition method to the nonlinear Boussinesq groundwater equation and the derived solution was tested with the observed data. \citet{Moutsopoulos2009} obtained an analytical solution for turbulent flow under the power-law inlet condition via Adomian's decomposition method. \citet{Olsen2019} expanded the problem under the turbulent flow condition studied by \citet{Moutsopoulos2009} by considering the polynomial approximate solutions for additional conditions such as the exponential head, power-law flux, and exponential flux conditions. \citet{Basha2021} derived the perturbation solution when the inlet condition is expressed as a linear function of positive time. 

Numerical methods, such as finite difference methods, finite element methods (FEM), and boundary element methods, are powerful tools for analyzing realistic boundaries \citep{Tang2006}. \citet{Taigbenu1991} proposed a Galerkin formulation of the Boussinesq groundwater equation by using linear basis functions. \citet{Cordano2013} provided a finite volume numerical solution for the two-dimensional Boussinesq equation, whose convergence was proven by \citet{Brugnano2008}. \citet{Tang2006} introduced a semianalytical time integration method by splitting the nonlinear term of the spatial discretization of the Boussinesq equation into a constant part and a variable part. 

In the current work, the problem of groundwater flow in semi-infinite soil with rapidly rising inlet water head conditions is expressed by the backward power law head condition. We perform the similarity transformation to the Boussinesq groundwater equation under this condition and obtain a nonlinear ordinary differential equation as well as the previous conditions \citep*{Barenblatt1990, Telyakovskiy2006}. The resulting ODE is numerically solved by applying Shampine's method \citep{Shampine1973}, and FEM is also performed for the original PDE. The comparison of both numerical computation shows the validity of similarity calculation. A quadratic approximate solution \citep{Lockington2000} is also derived by assuming the second-order power series expansion \citep{Song2007}.  Then, the approximate solution is compared with Shampine's solution.

\section{Initial and boundary value problem}

\begin{figure}[h]
		\centering
		\includegraphics{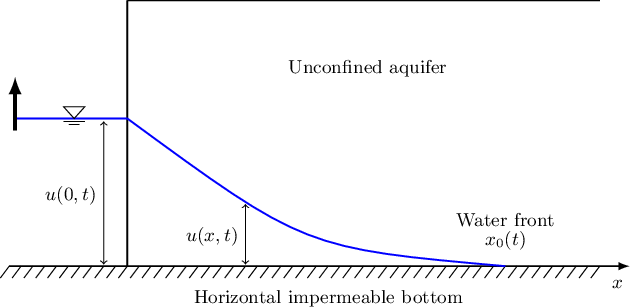}
		\caption{Schematic illustration of groundwater flow into an unconfined aquifer from an adjacent water source with a rapidly increasing boundary water level.}
		\label{fig:Illustration_aquifer}
	\end{figure}

In an unconfined aquifer, we introduce Dupuit's assumption that the flows are dominantly horizontal with hydrostatic pressure distribution. Then, integrating the conservation law of the saturated water zone with  Dupuit's assumption, the one-dimensional Boussinesq groundwater equation for subsurface water flows in unconfined aquifers is obtained as:
\begin{equation}
	S\frac{\partial u}{\partial t} = \frac{K}{2}\frac{\partial^2 u^2}{\partial x^2},  \label{eq:Boussinesq}
\end{equation}
where $S$ is the specific storage, $K$ is the saturated hydraulic conductivity, and $u $ is the elevation of the water table from the horizontal impermeable bottom, which is defined in $(x,t) = \mathbb{R}_+ \times (0, T)$. The groundwater equation (\ref{eq:Boussinesq}) is also valid for the power law hydraulic conductivity function when higher-order terms are negligible \citep{Furtak-Cole2018}. Then, we consider the groundwater flow into the unconfined aquifer with the boundary conditions
\begin{align}
	&u(0,t) = U(T-t)^{\alpha}, \label{InletCondition} \\
	&u(x_0, t) = 0, \label{SingularityCondition}
\end{align}
where $U>0$ and $\alpha \leq -1$ are constants, and where $x_0$ represents the wetting front position of the groundwater. A schematic illustration of the problem is shown in Figure \ref{fig:Illustration_aquifer}. 
Note that we assume that $u$ is a smooth non-negative classical solution for $0 \leq x \leq x_0$ with compact support. 
The initial condition is defined as
\begin{equation}
	u(x,0) = g_0(x), \label{t0Condition}
\end{equation}
where $g_0$ is a monotonically decreasing function starting from $UT^\alpha$ for $0 \leq x < x_0 (t)$, and vanishing for $x_0 (t) \leq x$, which is consistent with the similarity transformation described below. The explicit form is given in the last part of this section.

The inlet condition (\ref{InletCondition}) represents the water level of the adjacent water source, which increases exponentially with time and eventually diverges at $t = T$. Since the final time $T$ is set arbitrarily, it is possible to choose the time convenient for fitting objective hydraulic conditions. To date, the forward time condition $u(0,t) = U(t + T)^{\beta}$ has been considered an inlet condition in a wide range of literature \citep*{Barenblatt1990, Lockington2000, Telyakovskiy2002, Song2007}. Moreover, the exponential inlet condition $u(0,t) = U\exp(\gamma t)$ was also studied by \citet*{Barenblatt1990, Telyakovskiy2006} as the limiting case of the forward condition. In this context, the backward condition (\ref{InletCondition}) is seen as the situation beyond the exponential condition that is regarded as the singularity point (the visual implication is given below). 
		
With the use of Boltzmann's transformation under these conditions (\ref{InletCondition})-(\ref{t0Condition}), the governing equation (\ref{eq:Boussinesq}) can be reduced into the ordinary differential equation
\begin{equation}
	\frac{d^2H^2}{d\xi^{2}}-\frac{1}{2}(1+\alpha)\xi \frac{dH}{d\xi}+\alpha H = 0, \label{eq:BoussinesqODE}
\end{equation}
where we employ the dimensionless variable
\begin{align}
	&u = U(T-t)^{\alpha}H, \label{ScalingFunction} \\
	&\xi = x\sqrt{\frac{2S}{KU}}(T-t)^{-\frac{1+\alpha}{2}}. \label{DimensionlessVariable}
\end{align}
According to \citep{Vazquez2007}, the solution class does not change under the scaling (\ref{ScalingFunction})-(\ref{DimensionlessVariable}). 
The corresponding conditions are as follows:
\begin{align}
	&H(0) = 1, \label{InitialCondition} \\
	&H(\xi_0) = 0, \label{FrontCondition}\\
	&\frac{dH}{d\xi}(\xi_0^-) = \frac{1}{4} (1+\alpha) \xi_0. \label{FrontSlopeCondition}
\end{align}
The initial condition (\ref{InitialCondition}) arises from the inlet condition (\ref{InletCondition}) and the front condition (\ref{FrontCondition}) follows from the condition (\ref{SingularityCondition}). The existence of such a front $\xi_0$ is admitted by \citet{Gilding1976}, who prove the existence of a nontrivial similarity solution of the porous media equation with the backward inlet condition considered here. Furthermore, the front slope condition (\ref{FrontSlopeCondition}) is derived from the ordinary differential equation (\ref{eq:BoussinesqODE}) and the front condition (\ref{FrontCondition}) by considering the limiting situation $\xi \to \xi_0^-$, which is helpful for constructing the system for Shampine's method below.

Additionally, by using the solution of the boundary value problem for the ODE (\ref{eq:BoussinesqODE}), the initial condition for the original PDE (\ref{t0Condition}) is rewritten as
\begin{equation}
	u(x,0) = UT^{\alpha}H \left( x\sqrt{\frac{2S}{KU}}T^{-\frac{1+\alpha}{2}} \right).
\end{equation}

\section{Similarity solution and numerical solution}

\subsection{Approximate analytical solution}
In this section, we derive a quadratic approximate solution to the backward self-similar condition (\ref{InletCondition}). For the forward case, \citet{Lockington2000} first presented a quadratic approximate solution based on the analytical solution proposed by \citet{Barenblatt1990}, by retaining the first two terms of the corresponding power series expansion \citep{Song2007}.
To obtain a quadratic approximate solution of the nonlinear ordinary equation (\ref{eq:BoussinesqODE}) subject to the boundary conditions (\ref{InitialCondition})-(\ref{FrontCondition}), we assume a polynomial solution
\begin{equation}
	H(\xi) =
					\sum_{n=0}^{2}{a_n\left(1-\frac{\xi}{\xi_0}\right)^n}, \label{eq:PowerSeriesSol}
\end{equation}
where the solution must satisfy the boundary conditions
\begin{align}
	&H(0) =\sum_{n=0}^{2}{a_n} = 1, \label{InitialPS}\\
	&H(\xi_0) = a_0 = 0 \label{BoundaryPS}.
\end{align}
Substituting (\ref{eq:PowerSeriesSol}) into (\ref{eq:BoussinesqODE}) with the boundary condition (\ref{BoundaryPS}) yields the recurrence relation
\begin{equation}
	\frac{(n+1)(n+2)}{\xi_0^2}\left(\sum_{l=1}^{n+1}{a_l a_{n+2-l}}\right) 
	+ \frac{(1+\alpha)(n+1)}{2} a_{n+1}
	+ \left(\alpha- \frac{n}{2} (1+\alpha) \right)a_n = 0, \label{RecurrenceRelation}
\end{equation}
where $n = 0, 1$ and the signs of the last two terms differ from those of the previous term derived by \citet{Song2007}.
When $n = 0, 1$, the corresponding coefficients are
\begin{align}
	&a_1 = -\frac{\xi_0^2}{4}(1+\alpha), \label{eq:a1}\\
	&a_2 = -\frac{\alpha - 1}{16}\xi_0^2, \label{eq:a2}
\end{align}
where $a_1 = 0$ is disregarded as it lacks physical meaning. 
For $(n \geq 1)$, the recurrent relation can be rewritten explicitly as
\begin{equation}
	a_n = \frac{2(n+1)}{n(1+\alpha)\xi_0^2} \left( \sum_{l=2}^{n-1}{a_l a_{n+1-l}} \right) + \frac{(1+\alpha)(n+1) - 2\alpha}{n^2 (1+\alpha)} a_{n-1}.
\end{equation}
It should be noted that for $n = 3$, the coefficient $a_3$ is given by
\begin{equation}
	a_3 = \frac{(\alpha -1)(7-3\alpha)}{(1+\alpha)288} \xi_0^2.
\end{equation}
This expression diverges as $\alpha \to -1$. Consequently, we have focused our analysis on the cases $n = 0, 1, 2$.

These coefficients (\ref{eq:a1})-(\ref{eq:a2}) and the backward recurrence relation (\ref{RecurrenceRelation}) indicate that all the coefficients have the quadratic form, $a_n = b_n \xi_0^2$. Thus, the initial condition (\ref{InitialPS}) designates the front position 
\begin{equation}
 	\xi_0 = \left( \sum_{n=0}^{2}{b_n} \right)^{-\frac{1}{2}},\quad\text{where $b_n = \frac{a_n}{\xi_0^2}$}, \label{eq:FrontODE}
\end{equation}
which has the same form as the one presented by \citet*{Barenblatt1990,Song2007}.

\begin{figure}[h]
		\centering
		\includegraphics[width=80mm, height=60mm]{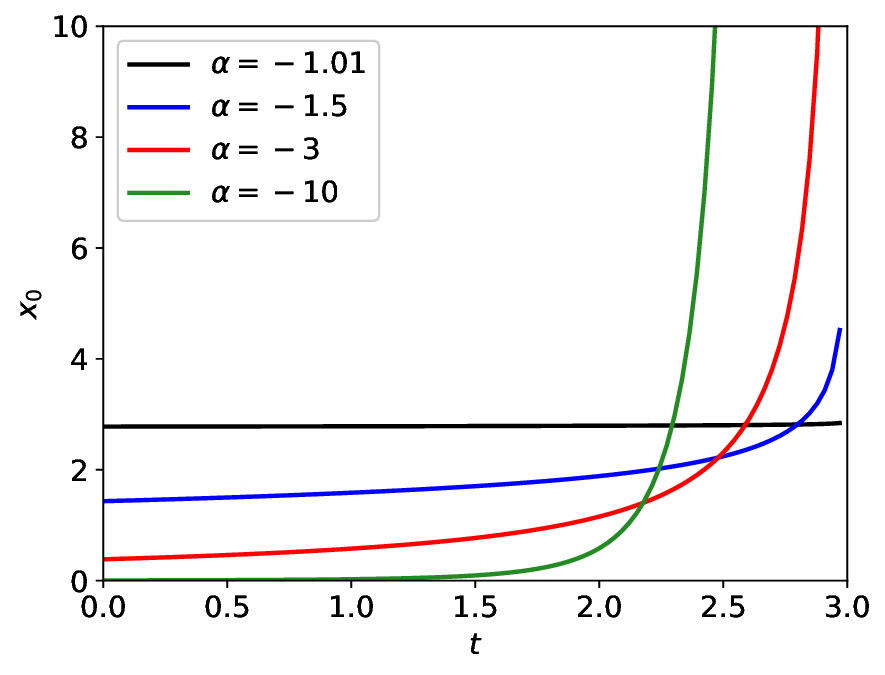}
		\caption{Time variation of the water front position $x_0$ for various $\alpha$ when $K=U =1$, $S=1/2$, and $T=3$ in the inlet condition (\ref{InletCondition}).}
		\label{fig:Backward_frontxt}
	\end{figure}

Then, we can construct a quadratic approximate solution 
\begin{equation}
	H(\xi) \approx -\frac{\xi_0^2}{4}(1+\alpha)\left(1-\frac{\xi}{\xi_0}\right) - \frac{\alpha - 1}{16}\xi_0^2\left(1-\frac{\xi}{\xi_0}\right)^2, \label{eq:QuadraticApproximate}
\end{equation}
where the front position is determined by
\begin{equation}
	\xi_0 \approx {\left(-\frac{1+\alpha}{4}-\frac{\alpha-1}{16}\right)}^{-\frac{1}{2}} 
		 = \left( - \frac{5\alpha+3}{16} \right)^{-\frac{1}{2}}. \label{eq:QuadraticFront}
\end{equation}
For the real scale, combining equations (\ref{DimensionlessVariable}) and (\ref{eq:QuadraticFront}) yields the water front $x_0(t)$,
\begin{equation}
	x_0(t) = \left( - \frac{2S(5\alpha + 3)}{16KU} \right)^{-\frac{1}{2}} (T - t)^{\frac{1+\alpha}{2}}. \label{eq:Waterfront}
\end{equation}
Figure \ref{fig:Backward_frontxt} represents the profile of the real front position $x_0$ for the Boussinesq equation (\ref{eq:Boussinesq}) with $K = U = 1$ and $S=1/2$, illustrating its temporal evolution for various $\alpha $.

\subsection{Shampine's method}
The derived quadratic approximate solutions (\ref{eq:QuadraticApproximate}) are compared with numerical solutions obtained using the method of \citet{Shampine1973}. 
This approach allows us to compute an auxiliary solution using the standard fourth-order Runge-Kutta method, after which the true solution is recovered through proper rescaling. 

\begin{figure}[h]
		\centering
		\includegraphics[width=80mm, height=60mm]{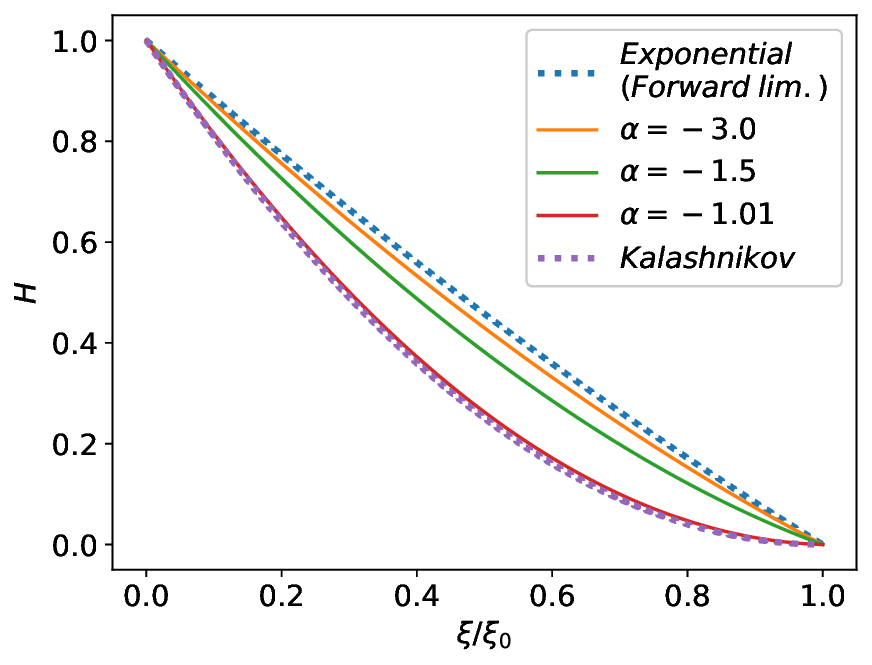}
		\caption{Numerical solutions of (\ref{eq:BoussinesqODE})-(\ref{FrontCondition}) for various $\alpha$ (solid line). Kalashnikov's solution for the lower limit and a numerical result of the exponential condition for the upper limit (dotted line) are also presented.}
		\label{fig:Backward_Shampine}
	\end{figure}

The original ODE (\ref{eq:BoussinesqODE}) is transformed into the following system
\begin{equation}
\left\{\,
	\begin{aligned}
		&\frac{df}{d\xi} = \frac{1}{4} (1+\alpha) \xi \frac{f}{H} - \frac{\alpha}{2}H,\\
		&\frac{dH}{d\xi} = \frac{f}{H}, \label{eq:BoussinesqSystem}
	\end{aligned}
\right.
\end{equation}
where we use the hodograph variable $f = HdH/d\xi$ known as a dimensionless total discharge \citep{Chen1995}. To perform the integration slightly away from $\xi = \xi_0$ where the first derivative of $H$ is discontinuous, we consider a Taylor series solution expanded about $\xi = \xi_0$ as initial conditions
\begin{equation}
	\begin{split}
	f(\xi_0) &\approx \frac{1}{4} (1 + \alpha) \xi_0 H(\xi_0),\\
	H(\xi_0) &\approx \frac{1}{4} (1+\alpha) \xi_0 (\xi - \xi_0) + \dotsb,
	\end{split}
\end{equation}
where the boundary conditions (\ref{FrontCondition}), (\ref{FrontSlopeCondition}) are employed.
Using a fourth-order Runge-Kutta method with a step of $1.0 \times 10^{-5}$, the first-order system (\ref{eq:BoussinesqSystem}) is integrated backward from $\xi = \xi_0 = 1$ to $\xi = 0$. The obtained numerical solution is the case when $\xi_0 = 1$ and may not satisfy the original problem (\ref{eq:BoussinesqODE})-(\ref{FrontCondition}). Owing to the nature of the similarity solution \citep{Gilding1976}, the results are scaled via the scaling formula
\begin{equation}
	H(\xi; \xi_0) = kH'(\sqrt{k}\xi; \sqrt{k}\xi_0) 
				= \frac{H'(\xi / \xi_0; 1)}{\xi_0^2}, \label{eq:ScalingFormula}
\end{equation} 
where $H'(\xi)$ denotes a rescaled solution derived under modified conditions, introduced here to represent the numerical results obtained via Shampine's method.
The scaling formula must satisfy the initial condition (\ref{InitialCondition}), and gives rise to the desired solution by calculating the front position
\begin{equation}
	\xi_0 = \sqrt{\frac{H'(0; 1)}{H(0; \xi_0)}}. \label{eq:CalFront}
\end{equation}
These numerical solutions for representative values of $\alpha$ are shown in Figure \ref{fig:Backward_Shampine} with solid lines.
When $\alpha = -1$, the obtained numerical solutions imply convergence to the known explicit solution 
\begin{equation}
	H_k(\xi) = 	\begin{cases}
					\left(1-\frac{\xi}{\xi_0}\right)^2 & \text{for \quad$0 \leq \xi < \xi_0$}  \\
					0 & \text{for \quad$\xi_0 \leq \xi$} \label{eq:Kalashnikov}
				\end{cases}
\end{equation} 
which is a dimensionless form of a porous medium solution presented in \citet{Kalashnikov1967}, following the similarity transformation discussed in \citet{Gilding1982}.
On the other hand, the numerical solution in the limit $\alpha \to -\infty$ has a good agreement with the classical exponential condition $u(0,t) = Uexp(\gamma t)$. This corresponds to the forward condition $u(0,t) = U(t + T)^{\beta}$ as $\beta \to \infty$  \citep*{Barenblatt1990, Telyakovskiy2002}. 
In fact, taking the limit $\alpha \to -\infty$ in the ODE (\ref{eq:BoussinesqODE}), the coefficient $(1+\alpha)/2$ is dominated by $\alpha/2$. Consequently, the governing equation reduces to
\begin{equation}
	\frac{d^2H^2}{d\xi^2} - \frac{\alpha}{2} \xi \frac{dH}{d\xi} + \alpha H = 0,
\end{equation}
which is consistent with the case of an exponential condition, and with the same form of the equation appearing in \citet{Gilding1976}. It is also possible to visualize groundwater intrusion on a real scale by combining the front calculation (\ref{eq:CalFront}) and the scaling (\ref{DimensionlessVariable}).

	\begin{figure}[h]
		\centering
		\includegraphics[width=80mm, height=60mm]{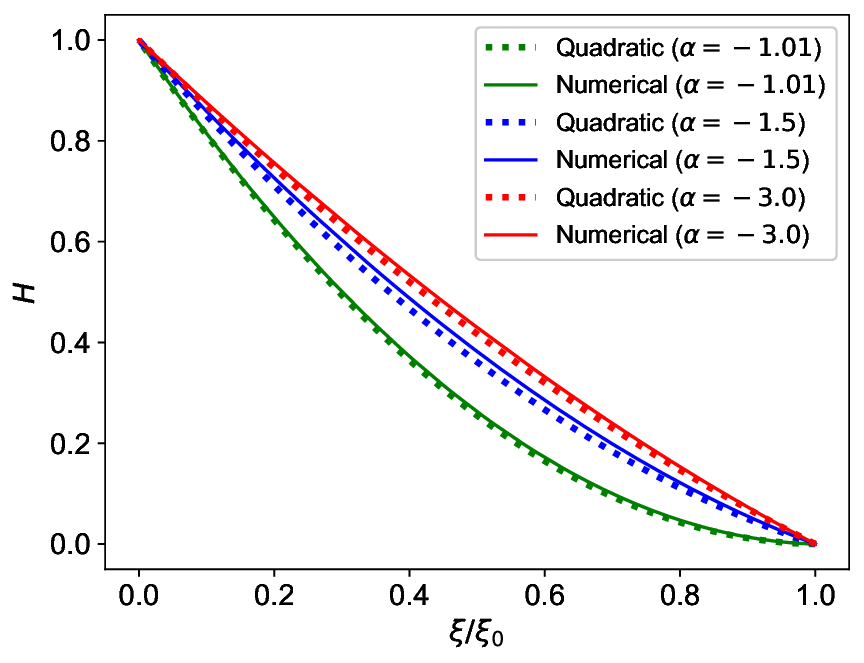}
		\caption{Comparison between the numerical solution and the approximate analytical solution.}
		\label{fig:Backward_song}
	\end{figure}

\begin{figure}[t]
		\centering
		\includegraphics[width=80mm, height=60mm]{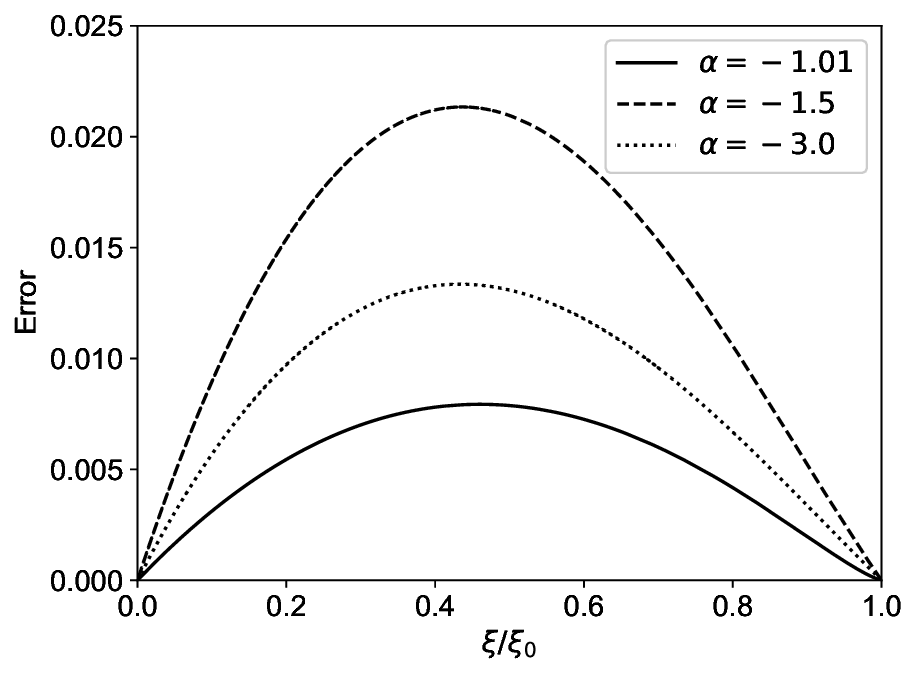}
		\caption{Errors defined by Shampine's numerical solution minus the quadratic approximate solution for various $\alpha$.}
		\label{fig:Backward_error}
	\end{figure}

Figure \ref{fig:Backward_song} shows the profiles of the numerical solutions via Shampine's method and the quadratic approximate solutions (\ref{eq:QuadraticApproximate}) with the quadratic front data (\ref{eq:QuadraticFront}) for different $\alpha$. The  relative error for the front $\xi_0$ is approximately 17\% for $\alpha = -1.01$, approximately 5\% for $\alpha  = -1.5$, and approximately 2\% for $\alpha = -3.0$. 
Figure \ref{fig:Backward_error} shows the error behavior for selected values of $\alpha$ ($-1.01$, $-1.5$, and $-3$). A decreasing trend in the error is observed as $\alpha$ approaches $-1$ and also as it moves toward more negative values (e.g., $\alpha=-3$), while relatively larger deviations occur at intermediate values such as $\alpha=-1.5$. This behavior indicates that the quadratic approximation becomes more accurate near the two limiting cases: the exponential case corresponding to $\alpha\to-\infty$, and the Kalashnikov case (\ref{eq:Kalashnikov}) corresponding to $\alpha\to-1$, where the approximation coincides with the analytical solution. 

These errors are larger than those of the forward condition written in \citet{Lockington2000} and \citet{Telyakovskiy2006} as a whole. The approximate expression (\ref{eq:QuadraticFront}) is fortunately admissible under the well-known forward time condition. On the other hand, the difference between the numerical value and the approximated value is greater with respect to the front position $\xi_0$ under the backward time condition considered above. Since the Barenblatt power series solution diverges in the case of the present backward condition because of its third coefficient unlike \citet{Song2007}, another type of analytical solution is needed to construct a more accurate approximate solution.

\subsection{Finite element method}
In the previous subsection, we employed Shampine's solutions as benchmark solutions. Herein, the accuracy of this solution is confirmed via a comparison with the solutions calculated via the simple finite element method. 
The Boussinesq groundwater equation (\ref{eq:Boussinesq}) can be written in a weak form
\begin{equation}
	\int_\Omega vS\frac{\partial u}{\partial t} dx = \int_\Omega v \frac{K}{2} \frac{\partial^2 u^2}{\partial x^2} dx 
												= -\int_\Omega \frac{K}{2} \frac{\partial v}{\partial x} \frac{\partial u^2}{\partial x} dx, \label{eq:WeakBoussinesq}
\end{equation}
where $\Omega \in \mathbb{R}$ is the computational domain, $v \in C_0^\infty(\Omega)$ is a test function, and $u$ is a weak solution. Then, the variables $v$ and $u$ are approximated by 
\begin{equation}
	v(x) \approx \sum^n_{i=0} v_i N_i (x), \quad u(x,t) \approx \sum^n_{i=0} u_i (t) N_i (x), \label{eq:FEMapprox}
\end{equation}
where $v_i$ are arbitrary constant coefficients; $u_i(t)$ are the nodal values of $u(x,t)$ at time $t$; $n$ is the number of total grids; and $N_i (x)$ is the piecewise linear basis function. The original PDE (\ref{eq:Boussinesq}) is transformed into the system of ODE
\begin{equation}
	\mathbf{A}\dot{\mathbf{u}} = -\mathbf{b}, \label{eq:FEMODE}
\end{equation}
with
\begin{align}
	A_{jk} &= \int_\Omega S N_j N_k dx, \\
	B_{jk} &= \int_\Omega \frac{K}{2} \frac{dN_j}{dx} \frac{dN_k}{dx} dx, \\
	b_{jk} &=  B_{jk} u^2_k,
\end{align}
where $j = 1, 2, \dotsb, n$, $k =0, 1, \dotsb, n$, and the value of $u^m_k$ is designated by the result of the previous step calculation. Additionally, imposing the same backward boundary condition (\ref{InletCondition}) identifies the value at the boundary grids
\begin{align}
	u_0 (t) &= U(T-t)^{\alpha}, \\
	A_{0k}  &= 
		\begin{cases}
			1 & \text{for } k = 0 \\
			0 & \text{for } k = 1, 2, \dotsb, n, 
		\end{cases}\\
	b_{0k}  &= 
		\begin{cases}
			u_0^2 & \text{for } k = 0 \\
			0 & \text{for } k = 1, 2, \dotsb, n.
		\end{cases}
\end{align}

\begin{figure}[h]
		\centering
		\includegraphics[width=80mm, height=60mm]{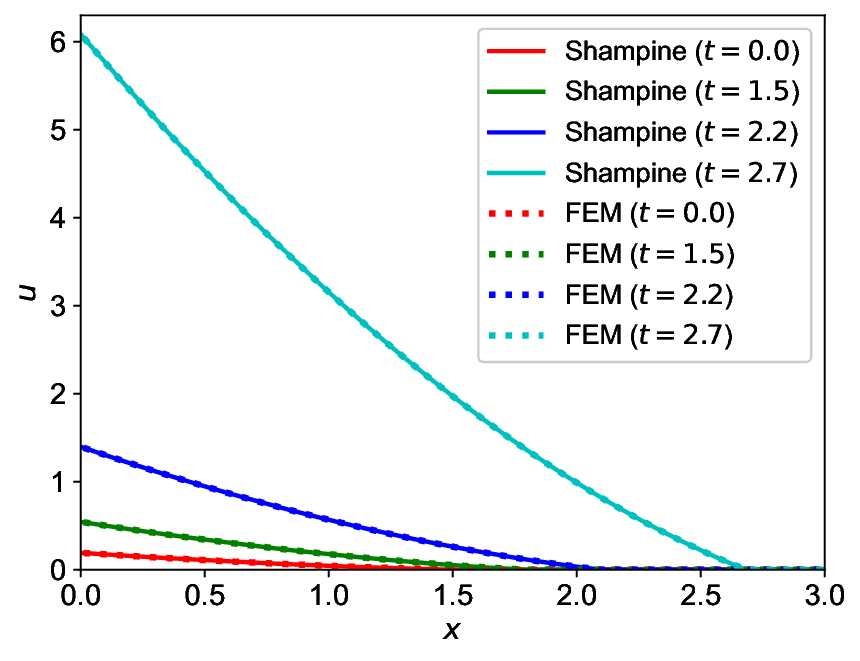}
		\caption{Real scale profile obtained from the interpolated Shampine's method (solid line), and the finite element method (dotted line) at times $t = 0.0, 1.5, 2.2, 2.7$.}
		\label{fig:Backward_real}
	\end{figure}

Herein, we construct the initial condition (\ref{t0Condition}) for the FEM computation from the numerical solution obtained by Shampine’s method, which provides solutions at arbitrary time instances with a much finer resolution than the FEM discretization; the required values are obtained by linear interpolation, as illustrated by the curve at $t = 0.0$ in Figure \ref{fig:Backward_real}.
	\begin{figure}[h]
		\centering
		\includegraphics[width=80mm, height=60mm]{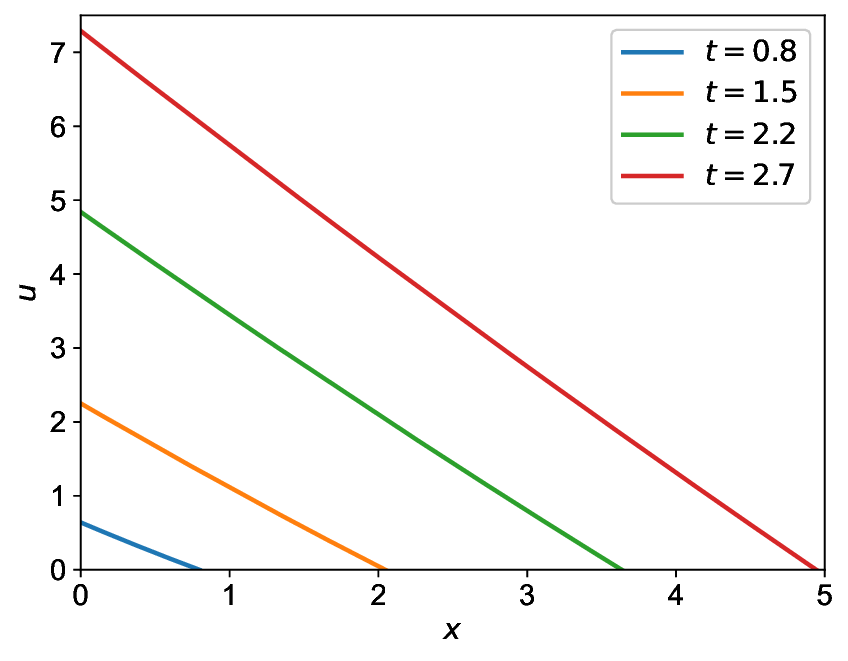}
		\caption{Real scale profile obtained from Shampine's method when the forward power-law head condition with $\beta = 2.0$ at times $t = 0.2, 0.4, 0.6, 0.8$.}
		\label{fig:Forward_real}
	\end{figure}
	\begin{figure*}
		\centering
		\includegraphics[width=\linewidth]{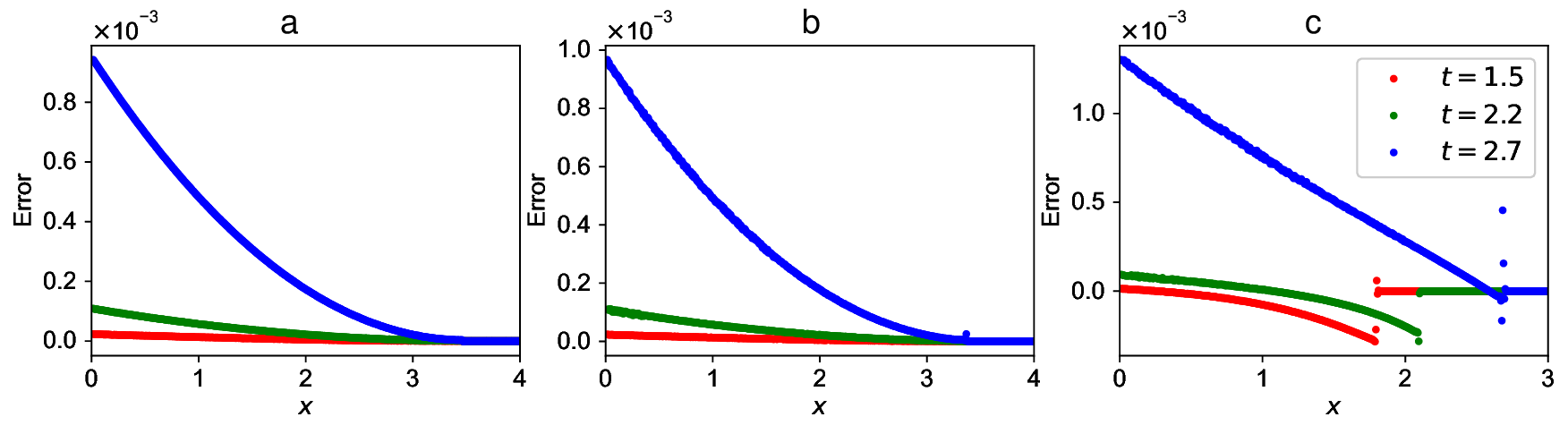}
		\caption{Errors of different numerical methods. Kalashnikov's solution minus FEM for $\alpha = - 1.0$ (a), and Shampine's solution minus FEM for $\alpha = -1.01$ (b), for $\alpha = -1.5$ (c).}
		\label{fig:FEM_error}
	\end{figure*}
The resulting spatial discretized equations with $\Delta x = 0.008$ are then calculated in the time direction via a fourth-order Runge-Kutta scheme with a step of $8.0 \times 10^{-7}$. Finally, the obtained algebraic equations are iteratively solved via the Gauss-Seidel method until the maximum absolute change in the values of $u_i$ becomes less than or equal to machine epsilon.

The numerical results are shown in Figure \ref{fig:Backward_real} where the solid line represents linear interpolated and scaled Shampine's solutions, and the dotted line illustrates the solution via the present FEM when $S = 1$, $K = 2$,  $U = 1$, $T = 3$, and $\alpha = -1.5$ for various times. The figure shows that the calculations via the FEM and Shampine's solution is in good agreement. It is observed that the solution is Lipschitz continuous as proved mathematically by \citet{Aronson1969} and the discontinuity of the first derivative becomes larger over time. The propagation profile of the present study differs from that of the previous papers \citep*{Lockington2000,Telyakovskiy2006} shown in Figure \ref{fig:Forward_real} where the same PDE of interest is transformed into ODE under the known initial and boundary conditions given as $u(x,0) = 0$, $u(0,t) = t^{2}$, and solved via Shampine's method. Since the rate of change at the boundary is much greater under the present backward power-law condition than under the classic forward time boundary condition, the water level increases before propagation reaches the waterfront.

The errors for different $\alpha$ values at times $t = 1.5, 2.2, 2.7$ are presented in Figure \ref{fig:FEM_error}. The error is defined by Shampine's solution minus the solution by FEM except when $\alpha = -1$ where it is possible to employ Kalashnikov's explicit solution (\ref{eq:Kalashnikov}) instead of Shampine's method.
The computational steps are summarized in Table \ref{tab:Numstep} where it is necessary for the time step width to be smaller than the space step width to satisfy the diffusion number requirement for numerical stability.
\begin{table}
	\caption{Computational step for different $\alpha$}
	\label{tab:Numstep}
	\begin{tabular}{ccc}
		\toprule
						& $\Delta x$ & $\Delta t$ \\ 
		\midrule
		$\alpha = -1.00$ & 0.008 & 0.0000008 \\
		$\alpha = -1.01$ & 0.008 & 0.0000008 \\
		$\alpha = -1.5$ & 0.006 & 0.0000006 \\
		\botrule
	\end{tabular}
\end{table}

In the case of Shampine, the error does not increase over time because of its similarity calculation characteristics, although the admissible conditions are limited to simple geometry as stated in Section \ref{intro}. The comparison between Figure \ref{fig:FEM_error}a and other plots implies the accuracy of Shampine's solution. The numerical results exhibit oscillatory behavior near the moving front, which is likely caused by the increasing non-smoothness of the solution at $\xi_0$ as $\alpha$ decreases.

\section{Conclusion}
The groundwater movement through a one-dimensional, unconfined aquifer was investigated under a power increasing inlet condition that diverges in finite time. This condition is thought to be an extension of the previous work \citep*{Barenblatt1990,Lockington2000,Telyakovskiy2006}, and enables the description of various groundwater infiltration situations owing to its flexibility of the final time $T$ in (\ref{InletCondition}) when the water level is blowing up at the boundary.
For these conditions, the Boussinesq groundwater equation is transformed into a nonlinear ordinary differential equation via newly introduced similarity transformation similar to the case of the forward time condition. Then, both of a quadratic approximate solution and numerical solutions via Shampine's method are derived as similarity solutions. The obtained quadratic solution approximated Shampine's numerical solution, although the front position calculated by the constructed recurrence relation (\ref{RecurrenceRelation}) diverges at $\alpha = -1$ when considering a higher-order approximation. Thus, a more general approximate analytical solution is desired. 

The upper limit of the obtained Shampine's solution agrees well with the limit of the previous forward condition and is therefore considered to be a natural extension of that one. On the other hand, for the lower limit, the calculation is in good agreement with Kalashnikov's solution of the porous media equation when the exponent $m = 2$. Then, the original PDE (\ref{eq:Boussinesq}) is numerically solved via the FEM which is considered a weak formulation of the equation. A comparison with the results of the finite element method provides Shampine's solution with a certain degree of validity as a benchmark, while the FEM has the advantage of being applicable to a wide range of real boundary conditions.

\bibliography{export}


\begin{thebibliography}{30}
\ifx \bisbn   \undefined \def \bisbn  #1{ISBN #1}\fi
\ifx \binits  \undefined \def \binits#1{#1}\fi
\ifx \bauthor  \undefined \def \bauthor#1{#1}\fi
\ifx \batitle  \undefined \def \batitle#1{#1}\fi
\ifx \bjtitle  \undefined \def \bjtitle#1{#1}\fi
\ifx \bvolume  \undefined \def \bvolume#1{\textbf{#1}}\fi
\ifx \byear  \undefined \def \byear#1{#1}\fi
\ifx \bissue  \undefined \def \bissue#1{#1}\fi
\ifx \bfpage  \undefined \def \bfpage#1{#1}\fi
\ifx \blpage  \undefined \def \blpage #1{#1}\fi
\ifx \burl  \undefined \def \burl#1{\textsf{#1}}\fi
\ifx \doiurl  \undefined \def \doiurl#1{\url{https://doi.org/#1}}\fi
\ifx \betal  \undefined \def \betal{\textit{et al.}}\fi
\ifx \binstitute  \undefined \def \binstitute#1{#1}\fi
\ifx \binstitutionaled  \undefined \def \binstitutionaled#1{#1}\fi
\ifx \bctitle  \undefined \def \bctitle#1{#1}\fi
\ifx \beditor  \undefined \def \beditor#1{#1}\fi
\ifx \bpublisher  \undefined \def \bpublisher#1{#1}\fi
\ifx \bbtitle  \undefined \def \bbtitle#1{#1}\fi
\ifx \bedition  \undefined \def \bedition#1{#1}\fi
\ifx \bseriesno  \undefined \def \bseriesno#1{#1}\fi
\ifx \blocation  \undefined \def \blocation#1{#1}\fi
\ifx \bsertitle  \undefined \def \bsertitle#1{#1}\fi
\ifx \bsnm \undefined \def \bsnm#1{#1}\fi
\ifx \bsuffix \undefined \def \bsuffix#1{#1}\fi
\ifx \bparticle \undefined \def \bparticle#1{#1}\fi
\ifx \barticle \undefined \def \barticle#1{#1}\fi
\bibcommenthead
\ifx \bconfdate \undefined \def \bconfdate #1{#1}\fi
\ifx \botherref \undefined \def \botherref #1{#1}\fi
\ifx \url \undefined \def \url#1{\textsf{#1}}\fi
\ifx \bchapter \undefined \def \bchapter#1{#1}\fi
\ifx \bbook \undefined \def \bbook#1{#1}\fi
\ifx \bcomment \undefined \def \bcomment#1{#1}\fi
\ifx \oauthor \undefined \def \oauthor#1{#1}\fi
\ifx \citeauthoryear \undefined \def \citeauthoryear#1{#1}\fi
\ifx \endbibitem  \undefined \def \endbibitem {}\fi
\ifx \bconflocation  \undefined \def \bconflocation#1{#1}\fi
\ifx \arxivurl  \undefined \def \arxivurl#1{\textsf{#1}}\fi
\csname PreBibitemsHook\endcsname

\bibitem[\protect\citeauthoryear{Aronson}{1969}]{Aronson1969}
\begin{barticle}
\bauthor{\bsnm{Aronson}, \binits{D.G.}}:
\batitle{Regularity properties of flows through porous media}.
\bjtitle{SIAM Journal on Applied Mathematics}
\bvolume{17},
\bfpage{461}--\blpage{467}
(\byear{1969})
\end{barticle}
\endbibitem

\bibitem[\protect\citeauthoryear{Basha}{2021}]{Basha2021}
\begin{barticle}
\bauthor{\bsnm{Basha}, \binits{H.A.}}:
\batitle{Perturbation solutions of the {Boussinesq} equation for horizontal
  flow in finite and semi-infinite aquifers}.
\bjtitle{Advances in Water Resources}
\bvolume{155},
\bfpage{104016}
(\byear{2021})
\doiurl{10.1016/j.advwatres.2021.104016}
\end{barticle}
\endbibitem

\bibitem[\protect\citeauthoryear{Brugnano and Casulli}{2008}]{Brugnano2008}
\begin{barticle}
\bauthor{\bsnm{Brugnano}, \binits{L.}},
\bauthor{\bsnm{Casulli}, \binits{V.}}:
\batitle{Iterative {Solution} of {Piecewise} {Linear} {Systems}}.
\bjtitle{SIAM Journal on Scientific Computing}
\bvolume{30}(\bissue{1}),
\bfpage{463}--\blpage{472}
(\byear{2008})
\doiurl{10.1137/070681867}
\end{barticle}
\endbibitem

\bibitem[\protect\citeauthoryear{Barenblatt et~al.}{1990}]{Barenblatt1990}
\begin{bbook}
\bauthor{\bsnm{Barenblatt}, \binits{G.I.}},
\bauthor{\bsnm{Entov}, \binits{V.M.}},
\bauthor{\bsnm{Ryzhik}, \binits{V.M.}}:
\bbtitle{Theory of Fluid Flows Through Natural Rocks}
vol. \bseriesno{395}.
\bpublisher{Kluwer Academic Publishers},
\blocation{Boston}
(\byear{1990})
\end{bbook}
\endbibitem

\bibitem[\protect\citeauthoryear{Brutsaert and Nieber}{1977}]{Brutsaert1977}
\begin{barticle}
\bauthor{\bsnm{Brutsaert}, \binits{W.}},
\bauthor{\bsnm{Nieber}, \binits{J.L.}}:
\batitle{Regionalized drought flow hydrographs from a mature glaciated
  plateau}.
\bjtitle{Water Resources Research}
\bvolume{13},
\bfpage{637}--\blpage{643}
(\byear{1977})
\doiurl{10.1029/WR013i003p00637}
\end{barticle}
\endbibitem

\bibitem[\protect\citeauthoryear{Chen et~al.}{1995}]{Chen1995}
\begin{barticle}
\bauthor{\bsnm{Chen}, \binits{Z.-X.}},
\bauthor{\bsnm{Bodvarsson}, \binits{G.S.}},
\bauthor{\bsnm{Witherspoon}, \binits{E.A.}},
\bauthor{\bsnm{Yortsos}, \binits{Y.C.}}:
\batitle{An integral equation formulation for the unconfined flow of
  groundwater with variable inlet conditions}.
\bjtitle{Transport in Porous Media}
\bvolume{18},
\bfpage{15}--\blpage{36}
(\byear{1995})
\end{barticle}
\endbibitem

\bibitem[\protect\citeauthoryear{Chor et~al.}{2013}]{Chor2013}
\begin{barticle}
\bauthor{\bsnm{Chor}, \binits{T.}},
\bauthor{\bsnm{Dias}, \binits{N.L.}},
\bauthor{\bsnm{Ruiz de Zárate}, \binits{A.}}:
\batitle{An exact series and improved numerical and approximate solutions for
  the {Boussinesq} equation}.
\bjtitle{Water Resources Research}
\bvolume{49},
\bfpage{7380}--\blpage{7387}
(\byear{2013})
\doiurl{10.1002/wrcr.20543}
\end{barticle}
\endbibitem

\bibitem[\protect\citeauthoryear{Cordano and Rigon}{2013}]{Cordano2013}
\begin{barticle}
\bauthor{\bsnm{Cordano}, \binits{E.}},
\bauthor{\bsnm{Rigon}, \binits{R.}}:
\batitle{A mass‐conservative method for the integration of the
  two‐dimensional groundwater ({Boussinesq}) equation}.
\bjtitle{Water Resources Research}
\bvolume{49}(\bissue{2}),
\bfpage{1058}--\blpage{1078}
(\byear{2013})
\doiurl{10.1002/wrcr.20072}
\end{barticle}
\endbibitem

\bibitem[\protect\citeauthoryear{Chor et~al.}{2019}]{Chor2019}
\begin{barticle}
\bauthor{\bsnm{Chor}, \binits{T.}},
\bauthor{\bsnm{Ruiz de Zárate}, \binits{A.}},
\bauthor{\bsnm{Dias}, \binits{N.L.}}:
\batitle{A generalized series solution for the {Boussinesq} equation with
  constant boundary conditions}.
\bjtitle{Water Resources Research}
\bvolume{55},
\bfpage{3567}--\blpage{3575}
(\byear{2019})
\doiurl{10.1029/2018WR024154}
\end{barticle}
\endbibitem

\bibitem[\protect\citeauthoryear{Furtak-Cole et~al.}{2018}]{Furtak-Cole2018}
\begin{barticle}
\bauthor{\bsnm{Furtak-Cole}, \binits{E.}},
\bauthor{\bsnm{Telyakovskiy}, \binits{A.S.}},
\bauthor{\bsnm{Cooper}, \binits{C.A.}}:
\batitle{A series solution for horizontal infiltration in an initially dry
  aquifer}.
\bjtitle{Advances in Water Resources}
\bvolume{116},
\bfpage{145}--\blpage{152}
(\byear{2018})
\doiurl{10.1016/j.advwatres.2018.04.005}
\end{barticle}
\endbibitem

\bibitem[\protect\citeauthoryear{Gilding}{1982}]{Gilding1982}
\begin{barticle}
\bauthor{\bsnm{Gilding}, \binits{B.H.}}:
\batitle{Similarity solutions of the porous media equation}.
\bjtitle{Journal of Hydrology}
\bvolume{56},
\bfpage{251}--\blpage{263}
(\byear{1982})
\end{barticle}
\endbibitem

\bibitem[\protect\citeauthoryear{Gilding and Peletier}{1976}]{Gilding1976}
\begin{barticle}
\bauthor{\bsnm{Gilding}, \binits{B.H.}},
\bauthor{\bsnm{Peletier}, \binits{L.A.}}:
\batitle{On a class of similarity solutions of the porous media equation}.
\bjtitle{Journal of Mathematical Analysis and Applications}
\bvolume{55},
\bfpage{351}--\blpage{364}
(\byear{1976})
\end{barticle}
\endbibitem

\bibitem[\protect\citeauthoryear{Hayek}{2024}]{Hayek2024}
\begin{botherref}
\oauthor{\bsnm{Hayek}, \binits{M.}}:
A simple and accurate closed-form analytical solution to the {Boussinesq}
  equation for horizontal flow
(2024)
\doiurl{10.1016/j.advwatres.2024.104628}
\end{botherref}
\endbibitem

\bibitem[\protect\citeauthoryear{Kalashnikov}{1967}]{Kalashnikov1967}
\begin{barticle}
\bauthor{\bsnm{Kalashnikov}, \binits{A.S.}}:
\batitle{The occurrence of singularities in solutions of the non-steady seepage
  equation}.
\bjtitle{USSR Computational Mathematics and Mathematical Physics}
\bvolume{7},
\bfpage{269}--\blpage{275}
(\byear{1967})
\doiurl{10.1016/0041-5553(67)90023-7}
\end{barticle}
\endbibitem

\bibitem[\protect\citeauthoryear{Kacimov and Šimůnek}{2021}]{Kacimov2021}
\begin{barticle}
\bauthor{\bsnm{Kacimov}, \binits{A.R.}},
\bauthor{\bsnm{Šimůnek}, \binits{J.}}:
\batitle{Analytical traveling-wave solutions and hydrus modeling of wet wedges
  propagating into dry soils: Barenblatt's regime for {Boussinesq's} equation
  generalized}.
\bjtitle{Journal of Hydrology}
\bvolume{598},
\bfpage{126413}
(\byear{2021})
\doiurl{10.1016/j.jhydrol.2021.126413}
\end{barticle}
\endbibitem

\bibitem[\protect\citeauthoryear{Lockington et~al.}{2000}]{Lockington2000}
\begin{barticle}
\bauthor{\bsnm{Lockington}, \binits{D.A.}},
\bauthor{\bsnm{Parlange}, \binits{J.-Y.}},
\bauthor{\bsnm{Parlange}, \binits{M.}},
\bauthor{\bsnm{Selker}, \binits{J.}}:
\batitle{Similarity solution of the {Boussinesq} equation}.
\bjtitle{Advances in Water Resources}
\bvolume{23}(\bissue{7}),
\bfpage{725}--\blpage{729}
(\byear{2000})
\end{barticle}
\endbibitem

\bibitem[\protect\citeauthoryear{Moutsopoulos}{2009}]{Moutsopoulos2009}
\begin{barticle}
\bauthor{\bsnm{Moutsopoulos}, \binits{K.N.}}:
\batitle{Exact and approximate analytical solutions for unsteady fully
  developed turbulent flow in porous media and fractures for time dependent
  boundary conditions}.
\bjtitle{Journal of Hydrology}
\bvolume{369},
\bfpage{78}--\blpage{89}
(\byear{2009})
\doiurl{10.1016/j.jhydrol.2009.02.025}
\end{barticle}
\endbibitem

\bibitem[\protect\citeauthoryear{Olsen et~al.}{2019}]{Olsen2019}
\begin{barticle}
\bauthor{\bsnm{Olsen}, \binits{J.S.}},
\bauthor{\bsnm{Mortensen}, \binits{J.}},
\bauthor{\bsnm{Telyakovskiy}, \binits{A.S.}}:
\batitle{Polynomial approximate solutions of an unconfined {Forchheimer}
  groundwater flow equation}.
\bjtitle{Advances in Water Resources}
\bvolume{123},
\bfpage{189}--\blpage{200}
(\byear{2019})
\doiurl{10.1016/j.advwatres.2018.11.002}
\end{barticle}
\endbibitem

\bibitem[\protect\citeauthoryear{Parlange et~al.}{2000}]{Parlange2000}
\begin{barticle}
\bauthor{\bsnm{Parlange}, \binits{J.-Y.}},
\bauthor{\bsnm{Hogarth}, \binits{W.L.}},
\bauthor{\bsnm{Govindaraju}, \binits{R.S.}},
\bauthor{\bsnm{Parlange}, \binits{M.B.}},
\bauthor{\bsnm{Lockington}, \binits{D.}}:
\batitle{On an exact analytical solution of the {Boussinesq} equation}.
\bjtitle{Transport in Porous Media}
\bvolume{39},
\bfpage{339}--\blpage{345}
(\byear{2000})
\end{barticle}
\endbibitem

\bibitem[\protect\citeauthoryear{Polubarinova-Kochina}{1962}]{Polubarinova1962}
\begin{bbook}
\bauthor{\bsnm{Polubarinova-Kochina}, \binits{P.Y.A.}}:
\bbtitle{Theory of Groundwater Movement}.
\bpublisher{Princeton University Press},
\blocation{Princeton}
(\byear{1962})
\end{bbook}
\endbibitem

\bibitem[\protect\citeauthoryear{Rupp and Selker}{2005}]{Rupp2005}
\begin{barticle}
\bauthor{\bsnm{Rupp}, \binits{D.E.}},
\bauthor{\bsnm{Selker}, \binits{J.S.}}:
\batitle{Drainage of a horizontal {Boussinesq} aquifer with a power law
  hydraulic conductivity profile}.
\bjtitle{Water Resources Research}
\bvolume{41},
\bfpage{1}--\blpage{8}
(\byear{2005})
\doiurl{10.1029/2005WR004241}
\end{barticle}
\endbibitem

\bibitem[\protect\citeauthoryear{Shampine}{1973}]{Shampine1973}
\begin{barticle}
\bauthor{\bsnm{Shampine}, \binits{L.F.}}:
\batitle{Some singular concentration dependent diffusion problems}.
\bjtitle{ZAMM Journal of Applied Mathematics and Mechanics}
\bvolume{53},
\bfpage{421}--\blpage{422}
(\byear{1973})
\doiurl{10.1002/zamm.19730530615}
\end{barticle}
\endbibitem

\bibitem[\protect\citeauthoryear{Song et~al.}{2007}]{Song2007}
\begin{barticle}
\bauthor{\bsnm{Song}, \binits{Z.-Y.}},
\bauthor{\bsnm{Li}, \binits{L.}},
\bauthor{\bsnm{Lockington}, \binits{D.}}:
\batitle{Note on {Barenblatt} power series solution to {Boussinesq} equation}.
\bjtitle{Applied Mathematics and Mechanics}
\bvolume{28}(\bissue{6}),
\bfpage{823}--\blpage{828}
(\byear{2007})
\doiurl{10.1007/s}
\end{barticle}
\endbibitem

\bibitem[\protect\citeauthoryear{Serrano and Workman}{1998}]{Serrano1998}
\begin{barticle}
\bauthor{\bsnm{Serrano}, \binits{S.E.}},
\bauthor{\bsnm{Workman}, \binits{S.R.}}:
\batitle{Modeling transient stream/aquifer interaction with the non-linear
  {Boussinesq} equation and its analytical solution}.
\bjtitle{Journal of Hydrology}
\bvolume{206}(\bissue{3-4}),
\bfpage{245}--\blpage{255}
(\byear{1998})
\end{barticle}
\endbibitem

\bibitem[\protect\citeauthoryear{Tang and Alshawabkeh}{2006a}]{Tang2006}
\begin{barticle}
\bauthor{\bsnm{Tang}, \binits{G.}},
\bauthor{\bsnm{Alshawabkeh}, \binits{A.N.}}:
\batitle{A semi-analytical time integration for numerical solution of
  {Boussinesq} equation}.
\bjtitle{Advances in Water Resources}
\bvolume{29}(\bissue{12}),
\bfpage{1953}--\blpage{1968}
(\byear{2006})
\doiurl{10.1016/j.advwatres.2006.02.003}
\end{barticle}
\endbibitem

\bibitem[\protect\citeauthoryear{Telyakovskiy and
  Allen}{2006b}]{Telyakovskiy2006}
\begin{barticle}
\bauthor{\bsnm{Telyakovskiy}, \binits{A.S.}},
\bauthor{\bsnm{Allen}, \binits{M.B.}}:
\batitle{Polynomial approximate solutions to the {Boussinesq} equation}.
\bjtitle{Advances in Water Resources}
\bvolume{29}(\bissue{12}),
\bfpage{1767}--\blpage{1779}
(\byear{2006})
\doiurl{10.1016/j.advwatres.2005.12.006}
\end{barticle}
\endbibitem

\bibitem[\protect\citeauthoryear{Taigbenu}{1991}]{Taigbenu1991}
\begin{barticle}
\bauthor{\bsnm{Taigbenu}, \binits{A.E.}}:
\batitle{A simplified finite element treatment of the nonlinear {Boussinesq}
  equation}.
\bjtitle{Advances in Water Resources}
\bvolume{14}(\bissue{1}),
\bfpage{42}--\blpage{49}
(\byear{1991})
\doiurl{10.1016/0309-1708(91)90029-N}
\end{barticle}
\endbibitem

\bibitem[\protect\citeauthoryear{Telyakovskiy et~al.}{2002}]{Telyakovskiy2002}
\begin{barticle}
\bauthor{\bsnm{Telyakovskiy}, \binits{A.S.}},
\bauthor{\bsnm{Braga}, \binits{G.A.}},
\bauthor{\bsnm{Furtado}, \binits{F.}}:
\batitle{Approximate similarity solutions to the {Boussinesq} equation}.
\bjtitle{Advances in Water Resources}
\bvolume{25},
\bfpage{191}--\blpage{194}
(\byear{2002})
\end{barticle}
\endbibitem

\bibitem[\protect\citeauthoryear{Tolikas et~al.}{1984}]{Tolikas1984}
\begin{barticle}
\bauthor{\bsnm{Tolikas}, \binits{P.K.}},
\bauthor{\bsnm{Sidiropoulos}, \binits{E.G.}},
\bauthor{\bsnm{Tzimopoulos}, \binits{C.D.}}:
\batitle{A simple analytical solution for the {Boussinesq} one‐dimensional
  groundwater flow equation}.
\bjtitle{Water Resources Research}
\bvolume{20},
\bfpage{24}--\blpage{28}
(\byear{1984})
\doiurl{10.1029/WR020i001p00024}
\end{barticle}
\endbibitem

\bibitem[\protect\citeauthoryear{Vazquez}{2007}]{Vazquez2007}
\begin{bbook}
\bauthor{\bsnm{Vazquez}, \binits{J.L.}}:
\bbtitle{The Porous Medium Equation: Mathematical Theory}.
\bpublisher{Oxford University Press},
\blocation{New York}
(\byear{2007})
\end{bbook}
\endbibitem

\end{thebibliography}

\section*{Statements \& Declarations}
\begin{itemize}
\item Funding

This work was supported by Grant-in-Aid for Scientific Research No.19KK0167 from the Japan Society for the Promotion of Science (JSPS), Research Promotion Support 2024A from the Kyoto University Foundation.

\item Competing interests 

The authors have no competing interests to declare that are relevant to the content of this article.

\item Author contribution

All authors contributed to the study conception, design, data collection and analysis. The first draft of the manuscript was written by Shuntaro Togo and Koichi Unami commented on previous versions of the manuscript. All authors read and approved the final manuscript.

\item Data availability 

The datasets generated during and analysed during the current study are available from the corresponding author on reasonable request.
\end{itemize}

\end{document}